\documentstyle[11pt]{article}
 
\def\pa{\partial}
 
\def\g{\gamma} \def\G{\Gamma}
\def\a{\alpha} 
\def\b{\beta} 
\def\d{\delta} 
\def\e{\epsilon}

\def\l{\lambda} \def\L{\Lambda}
\def\m{\mu} 
\def\n{\nu}

\def\s{\sigma}

\def\be{\begin{equation}}
\def\ee{\end{equation}}
     
\begin{document}

\begin{flushright} 
BRX TH--449\\
CTP\#2818
\end{flushright}
\begin{center}
{\large\bf Higher Derivative Chern--Simons Extensions}

\vspace{.2in}

S. Deser \\
Department of Physics, Brandeis University \\
Waltham, MA 02454, USA

and

R. Jackiw \\
Department of Physics, Massachusetts Institute of Technology \\
Cambridge, MA 02139, USA
\end{center}

We study the higher-derivative extensions of the D=3 Abelian
Chern--Simons topological invariant that would appear in a
perturbative effective action's momentum expansion.  
The leading, third-derivative, extension 
$I_{\scriptscriptstyle ECS}$ turns out to be
unique.  It remains parity-odd but 
depends only on the field strength, hence no
longer carries large gauge information, nor is it topological
because metric dependence accompanies the additional 
covariant derivatives, whose positions are seen to be 
fixed by gauge invariance.
Viewed as an independent action, $I_{\scriptscriptstyle ECS}$
requires the field strength to obey
the wave equation.  The more interesting model, 
adjoining $I_{\scriptscriptstyle ECS}$ 
to the Maxwell action, 
describes a pair of excitations. One is massless, 
the other a massive ghost, as
we exhibit both via the propagator and by performing the 
Hamiltonian decomposition.  We also present this model's
total stress tensor and energy.  Other actions
involving  $I_{\scriptscriptstyle ECS}$ are also noted.

\noindent{\bf 1. Introduction}

The remarkable properties of the Chern--Simons (CS) topological
invariant in D=3 gauge theories \cite{001,002} are by now
well-appreciated.  For Abelian vector fields, 
$I_{\scriptscriptstyle CS} =  m / {\scriptstyle 2} \, \int d^3x \, 
\e^{\a\b\g} A_\a \pa_\b A_\g$
is parity violating, of first derivative order, metric-independent,
and gauge invariant.  Viewed as an action, 
$I_{\scriptscriptstyle CS}$ leads
to the locally ``flat" field equation
$F_{\a\b} \equiv \pa_\a A_\b - \pa_\b A_\a = 0$.
Instead, when the Maxwell action 
$I_{\scriptscriptstyle MAX}$ is adjoined, the resulting
topologically massive electrodynamics (TME) describes a 
helicity $\pm$1 (depending on the sign of the
mass $m$) mode \cite{002}.  
The gravitational  CS analog is of third derivative
order  $(I_{\scriptscriptstyle CS} \sim 
m^{-1} \int \G R )$, and its
Euler--Lagrange equation 
$C^{\m\n} \equiv \e^{\m\a\b} D_\a (R^\n_\b - \frac{1}{4} \d^\n_\b 
R ) = 0$
states that space is locally conformally 
flat: the Cotton tensor $C^{\m\n}$ is the 3-space conformal 
tensor.
Adjoining the Einstein action leads to a dynamical system,
topologically massive gravity (TMG); despite its higher
order, the linearized limit of TMG describes 
a massive helicity $\pm$2 excitation \cite{002}.

The first derivative $I_{\scriptscriptstyle CS}$ appears
naturally in a perturbative
effective action expansion of QED$_{3}$, because the electron's 
mass term is
also P-violating in D=3, and higher derivative extensions of it
(which we denote by $I_{\scriptscriptstyle ECS}$) 
should appear as well \cite{003}
in a $(\pa/m)$ power series,
$$
I_{\scriptscriptstyle EFF} [A_\m ] = I_{\scriptscriptstyle CS} + 
I_{\scriptscriptstyle MAX} + I_{\scriptscriptstyle ECS} + {\cal O} 
(m^{-2}) \;.
$$
It is therefore a natural
question, both in connection with this expansion and for
comparison with the third derivative TMG, to consider such
$I_{\scriptscriptstyle ECS}$.  Additional derivative 
powers must be even (if
the parity violating $\e^{\a\b\g}$ is retained), the lowest
 extensions being the most illuminating.
In actuality, there is only one such extension, 
as exhibited by the equalities
\begin{eqnarray}
I_{\scriptscriptstyle ECS} & = & (2m)^{-1} \int d^3x 
\: \e^{\a\b\g}
\Box A_\a \pa_\b A_\g \; = \; - (2m)^{-1} \int
d^3x \e^{\a\b\g} \pa_\l A_\a \pa^\l \pa_\b A_\g \nonumber \\
& = & - (2m)^{-1} \int d^3x \: \e^{\a\b\g} f_\a\pa_\b f_\g \; ,
\;\;\; f^\a \equiv \textstyle{\frac{1}{2}} \: \e^{\a\m\n}
F_{\m\n} \;  .
\end{eqnarray}
Each term follows from its predecessor by an obvious 
integration by parts.  Hence, unlike the original 
$I_{\scriptscriptstyle CS}$, $I_{\scriptscriptstyle ECS}$
depends locally on the field strength and {\it not} on the 
potential, and so carries no ``large gauge" information.  
Our signature 
conventions are (+,--,--),
$\e^{012} = + 1 = \e_{012}$.

We shall firstly exhibit the excitations described by actions
containing $I_{\scriptscriptstyle ECS}$, including especially 
its sum with the Maxwell action
(ETME). There, we find a massless particle plus a massive ghost.
We then consider $I_{\scriptscriptstyle ECS}$ in a 
gravitational background,
where the higher derivatives necessarily engender metric 
dependence,
in contrast to the topological character 
of $I_{\scriptscriptstyle CS}$, and give rise to a stress tensor
that contributes explicitly to the 
energy of ETME, as we shall display in terms of the two
degrees of freedom.

\noindent{\bf 2. Maxwell--ECS Dynamics}

We shall work in source-free flat space in this section,
our aim being to characterize
the excitations described by actions that include
$I_{\scriptscriptstyle ECS}$.
It is clear from (1) that, taken alone, 
$I_{\scriptscriptstyle ECS}$ 
yields the unconstrained
massless propagation of the field strength,
\be
\Box f^\m = 0 \; .
\ee
Next, define the extended system by adjoining
 to $I_{\scriptscriptstyle ECS}$ the Maxwell action
$$
I_{\scriptscriptstyle ETME} =  - \textstyle{\frac{1}{2}} 
\int d^3x 
[ f^2_\m + 
m^{-1} \e^{\a\b\g} f_\a \pa_\b f_\g ] \eqno{(3{\rm a})}
$$
resulting in the field equations
$$
m \: \d I_{\scriptscriptstyle ETME}/\d A_\m   
= \Box f^\m - m \e^{\m\a\b} \pa_\a f_\b 
= - \e^{\m\a\b} \pa_\a (mf_\b + \e_\b~\!\!^{\g\d}
\;  \pa_\g f_\d )
= 0 \; .
\eqno{(3{\rm b})}
$$
\renewcommand{\theequation}{\arabic{equation}}
\setcounter{equation}{3}

\noindent Here $m$ is seen to have dimensions of inverse 
length or mass.
[We remark that (3a) is known \cite{004} to be  precisely 
equivalent to
TME, the Maxwell--CS action, if $f^\m$ is taken to be the 
fundamental variable rather than, as for us, the curl of 
an underlying
vector potential. Our equations are correspondingly 
the curl
of those of TME, as shown by the last equality in (3b).]
Note that 
ETME can be formally obtained from (the appropriate 
helicity branch
of) TME 
by the replacement $m\rightarrow
\Box /m$.  Hence, we can immediately write the
form of our propagator, using that of TME.  There \cite{002},
\be
G^{\m\n}_{\scriptscriptstyle TME} = (\Box + m^2)^{-1}
(g^{\m\n} -  (m/\Box ) \e^{\m\a\n} \pa_\a ) 
\ee
when acting on conserved sources; it clearly described 
a massive
excitation.  Hence, the ETME propagator becomes
\be
G^{\m\n}_{\scriptscriptstyle ETME} = 
(\Box + \Box^2/m^2)^{-1} (g^{\m\n} - m^{-1}
\e^{\m\a\n} \pa_\a ) \; ,
\ee
as of course follows directly from (3a).
This time, the denominator describes two excitations,
\be
m^2 \Box^{-1} (\Box + m^2)^{-1} =
\Box^{-1} - (\Box + m^2)^{-1} \; .
\ee
One is massless, the other is massive, with a relative 
ghost sign.  The
limit $m\rightarrow\infty$ of 
$G_{\scriptscriptstyle ETME}$
correctly reproduces the Maxwell propagator
$g^{\m\n} /\Box$.  Its small $m$ limit should correspond 
to pure
$I_{\scriptscriptstyle ECS}$ and indeed we find 
$G_{\scriptscriptstyle ETME} \rightarrow - m/\Box^2 \; 
\e^{\m\a\n}
\pa_\a$, the propagator corresponding to (2). Note that,
irrespective of the signs in (3), there is never a tachyon:
there cannot be a $(\Box -m^2 )$ pole. 

We next perform a detailed canonical analysis, decomposing
$f^\m$ in terms of the vector potential ($A_i ,\; A_0$),
and writing
\be
A_i  \equiv \e_i~^j \hat{\pa}_j a + \pa_i \L \; , \;\;\;\;\;\;
\hat{\pa}_i \equiv \pa_i / \sqrt{-{\scriptstyle \nabla^2}}
\; , \;\;\;\;\;\;
(\e_i~^j \equiv -\e^{ij}) \; ,
\ee
to yield
\be
f^0 = \e^{ij}\pa_i A_j  = - \sqrt{-{\scriptstyle \nabla^2}} 
\; a \; , \;\;\;\;
f_i = \hat{\pa}_i \dot{a} + \e_{ij} \hat{\pa}_j \; E\; ,
\;\;\;\;\;\;\;
E \equiv - \sqrt{-{\scriptstyle \nabla^2}} (A_0 -\dot{\L}) \; .
\ee
The action $I_{\scriptscriptstyle ETME}$ then reduces to
\be
I_{\scriptscriptstyle ETME} = \textstyle{\frac{1}{2}}
\, \int d^3x 
(-a\Box a + E^2 ) + m^{-1} \int d^3x
E\Box a \; .
\ee
The Maxwell contribution $(m=\infty )$ is that of the 
transverse $a$-mode,
together with a nonpropagating longitudinal electric 
field term.
Note the absence of dangerous explicit third time 
derivative terms.  This is 
unsurprising: they can only come from 
the $\int \e^{ij} f_i \pa_0 f_j$ part of 
$I_{\scriptscriptstyle ECS}$.
Due to the explicit $\e^{ij}$,  all ``diagonal" terms
($a a$) and ($EE$) vanish by antisymmetry
(after spatial partial integration) 
since these scalars would have to carry
$\pa_i, \pa_j$ to saturate $\e^{ij}$.
The 
$aE$ cross-term has no $\e^{ij}$ and so can and does 
contain the third
derivative 
$a \: \pa^3_0 \, 
\L$, but (since $\L$ is a gauge parameter)
one $\pa_0$ is harmlessly buried as part of
the gauge-invariant field variable $E$.
The field equations from (9) are obviously
\be
\Box (a -  \bar{E}) = 0 \; , \;\;\;
\Box a + m^2 \bar{E} = 0 \; , \;\;\;\;\;\;\;
\bar{E} \equiv m^{-1} E \; .
\ee
The appropriate diagonalization is also clear from (9):
\be
I_{\scriptscriptstyle ETME} = - \textstyle{\frac{1}{2}} 
\int d^3x \;
\bar{a} \Box \bar{a} +  \textstyle{\frac{1}{2}} 
\int d^3x \; \bar{E} (\Box + m^2 )\bar{E}
\ee
in terms of $\bar{a} \equiv a - \bar{E}$.  Here we
have the normal transverse massless photon $\bar{a}$, 
together with a massive ghost $\bar{E}$ (longitudinal 
electric field), as
evidenced by the relative minus sign between the two 
modes.  We  again
recover the Maxwell theory in the $m\rightarrow\infty$ limit,
while as $m\rightarrow 0$, we find 
$\Box f^\m = 0$ in terms of the two invariant 
parts $(a,E)$ of $f^\m$.  All this agrees with the analysis
above of the propagator (5,6).  We have not studied the 
spin character of our excitations; any massless particle
must be spinless in D=3
\cite{005}, while the massive mode presumably has
helicity $\pm 1$ according to the sign of $m$.
We also omit details of coupling {ETME} 
to sources,
which is a straightforward exercise in the propagator's
properties for minimally coupled $\sim A_\m j^\m$
currents.  Non-minimal interactions $\sim f_\m k^\m$
are also permitted here, and would presumably be
equivalent to minimal couplings in TME, in view of
the equivalence of (3) to TME, in terms of
$f_\m$ alone.

The most general gauge invariant action involving the
terms discussed so far (apart from obvious $\Box^2$
or higher
insertions in $I_{\scriptscriptstyle CS}$) 
would be the linear combination
\be
I_{\scriptscriptstyle TOT} = I_{\scriptscriptstyle ECS} 
+ I_{\scriptscriptstyle MAX} + I_{\scriptscriptstyle CS} \; .
\ee
The only unexplored 2-term action, $I_{\scriptscriptstyle ECS} 
+ I_{\scriptscriptstyle CS}$,
obviously just corresponds to massive propagation,
$(\Box + m^2 ) f^\m = 0$, of the field strength.
If we keep all three terms, the propagator can once
again be read off from $G^{\m\n}_{\scriptscriptstyle TME}$ by the 
replacement $m \rightarrow m + c\Box /m$, where we
have allowed for different mass parameters
$m, \; m/c$ in the
two CS variants.  The $G_{\scriptscriptstyle TOT}$ 
denominator has the form
$m^{-2} [c^2 \Box^2 + m^2 \Box (1+2c) + m^4]$
with roots $m^2[-(1+2c) \pm (1+4c)^{1/2}] / 2c^2$.
The degenerate root at $c = -\frac{1}{4}$ corresponds
to a double (massive) pole.  Clearly there is a
range of special cases, both physical and not, that can
be explored, but no massless excitation remains.

\noindent{\bf 3. Curved Space Gauge Invariance}

When the geometry is nonflat, even the Maxwell action
must be written properly to preserve gauge invariance.
For example if instead of writing the manifestly
invariant form $F_{\m\n} F^{\m\n}$, we had continued
from its equally correct flat space expansion
$(\pa_\m A_\n)^2-(\pa^\m A_\m)^2$ to the covariant expression
$(D_\m A_\n)^2-(D^\m A_\m)^2$, gauge invariance would
be lost; the only correct order in the last term 
is $(D_\m A_\n)(D^\n A^\m)$, which differs
\cite{006} from $(D^\m A_\m )^2$ by a gauge-variant,
curvature-dependent term
$R^{\m\n}A_\m A_\n$.  It is similarly easy to see
that only the last, manifestly
gauge invariant derivative ordering of
(1) preserves curved space gauge
invariance
\renewcommand{\theequation}{\arabic{equation}}
\setcounter{equation}{12}
\be
I_{ECS} = -(2m)^{-1} \int \, d^3x \, \e^{\a\b\g}
f_\a \pa_\b f_\g \; , \;\;\;
f_\a \equiv  g^{-1/2} g_{\a\b}
\e^{\b\m\n} \pa_\m A_\n   \; .
\ee
Note that as defined here, $f_\a$ is a covariant vector; 
$f^a$ will denote
the contravariant vector (not the usual density)
$g^{\a\b}f_\b$.
The metric dependence $I_{\scriptscriptstyle ECS}$
is now entirely contained in
$f_\a$, so that the stress tensor is 
\be
 \sqrt{g} \: T^{\m\n}_{\scriptscriptstyle ECS} =
2 \d I_{\scriptscriptstyle ECS}/\d g_{\m\n} =  
-m^{-1} \left\{ (\e^{\m\a\b} f^\n + \e^{\n\a\b} f^\m ) \pa_\a
f_\b - g^{\m\n} \e^{\a\b\g} f_\a \pa_\b f_\g \right\} \; ,
\ee
in contrast to $T^{\m\n}_{\scriptscriptstyle CS}\equiv
0$.
The (covariant) conservation of $T^{\m\n}$ on ECS
shell is easily checked, using the Bianchi identities
$\pa_\m (\sqrt{g} \, f^\m ) \equiv 0$, 
the (covariant) field equations
\be
\e_\m~^{\a\l} \: \pa_\l (g^{-1/2} \e_\a~^{\b\g} \pa_\b f_\g )
\equiv D^\n (-\pa_\m f_\n + \pa_\n f_\m )
\equiv [g_{\m\n} \, D^2 - D_\n D_\m ]
f^\n = D^2 f^\m - R^{\m\n} f_\n =  0
\ee
and the identity $\e^{\a\b\g} \pa_\b f_\g
(\pa_\a f_\m - \pa_\m f_\a ) \equiv 0$.

The total ECSE stress tensor includes the usual
Maxwell contribution,
$$
T^{\m\n}_{\scriptscriptstyle ECSE} = -\textstyle{\frac{1}{2}}
(f^\m f^\n + f^\n f^\m - 
g^{\m\n} f^\l f^\s g_{\l\s} )
+ T^{\m\n}_{\scriptscriptstyle ECS} \eqno{(16{\rm a})}
$$
and is likewise conserved on combined shell, 
where is can be written very simply: By virtue of the middle
term in (3b),  $T^{\m\n}_{\scriptscriptstyle ECSE}$
reduces to its Maxwell part but (in flat space)
with the operator
$(1+2m^{-2} \Box )$ rather than just unity, between the
$f$'s:
$$
-2 m^2 \, T^{\m\n}_{\scriptscriptstyle ECSE} =
f^\m (m^2 + 2\Box ) f^\n +
f^\n (m^2 + 2\Box ) f^\m
 - \, \eta^{\m\n} f^\a (m^2 + 2\Box ) f_{\a} \; .
\eqno{(16{\rm b})}
$$
[In curved space, $g_{\a\b} \Box $ is replaced
by the operator   $(D^2
g_{\a\b} - R_{\a\b})$ of (15).]  Inserting the
canonical decomposition (7,8) into 
$T^{00}_{\scriptscriptstyle ECSE}$ directly gives
the total energy as the expected difference between
photon and ghost mode contributions; in our signature,
\renewcommand{\theequation}{\arabic{equation}}
\setcounter{equation}{16}
\begin{eqnarray}
P_0 & = & - \int d^2 r \, T^{00} =  
\textstyle{\frac{1}{2}} m^{-2} \int d^2r 
[f_i (m^2 + 2\Box )f_i + f_0 (m^2 + 2\Box )f_0 ] \nonumber \\
&& = \textstyle{\frac{1}{2}} \, \int d^2r 
\left\{
\left[ (\pa_0\bar{a} )^2 + (\mbox{\boldmath$\nabla$}
 \bar{a})^2 
\right] -
\left[ (\pa_0\bar{E})^2 + (\mbox{\boldmath$\nabla$}
 \bar{E} )^2 + m^2
\bar{E}^2 \right]\right\} \; .
\end{eqnarray}

\noindent{\bf 4. Summary}

We have studied the first higher derivative analog
of the CS topological invariant, which would arise
in the effective QED$_3$ action's
expansion in powers of $\pa/m$.  This 
$I_{\scriptscriptstyle ECS}$ invariant turns out
to be unique, and while formally similar to 
$I_{\scriptscriptstyle CS}$, differs profoundly 
from it in two respects:  
first, $I_{\scriptscriptstyle ECS}$ is a local
function of the field strength, insensitive
to the ``large gauge" aspects captured by 
$I_{\scriptscriptstyle CS}$; second, it is no
longer topological but depends explicitly on the
background geometry. When $I_{\scriptscriptstyle ECS}$ 
is added to the Maxwell action, the resulting ETME system 
describes two degrees of
freedom, one massless, the other a massive ghost.  This
is in contrast with the otherwise similar gravitational
TMG model:
while both are of overall third (but of
second time) derivative order, TMG represents a
single massive excitation.  The reasons for this
difference can be traced to the roles played by the
respective component actions: First,
the Maxwell term, unlike the
Einstein one in TMG, already describes a 
(massless) degree of freedom. Second, the triple
derivative  CS term in gravity is conformally invariant and
this higher symmetry, absent in $I_{\scriptscriptstyle ESC}$,
eliminates one candidate mode.

This work was supported by NSF grant PHY93--18511 and
DOE grant DE--FC02--94ER40818.

\end{document}